\documentclass[prl, twocolumn, showpacs, nofootinbib, amsmath,amssymb, floatfix, eqsecnum]{revtex4-1}
\pdfoutput=1
\usepackage{amsmath}
\usepackage{amssymb}
\usepackage{amsthm}
\usepackage{amsfonts}
\usepackage{comment}
\usepackage[normalem]{ulem}
\usepackage{graphicx}
\usepackage{color,framed}
\usepackage{hyperref}
\usepackage{times}
\usepackage{enumerate}
\usepackage{lipsum}
\usepackage{slashed}
\usepackage{url}
\usepackage{bbm}
\usepackage{chngcntr}
\usepackage{tikz,pgfplots}
\counterwithout{equation}{section}
\usepackage{tikz}
\pgfplotsset{compat=1.18}

\hypersetup{
    colorlinks=true, 
    linktoc=all,     
    linkcolor=blue,  
}

\def \beq {\begin{equation}}
\def \eeq {\end{equation}}
\def \beqa {\begin{eqnarray}}
\def \eeqa {\end{eqnarray}}
\def \bseq {\begin{subequations}}
\def \eseq {\end{subequations}}

\newcommand \tr {{\rm tr}\,}
\newcommand \Tr {{\rm Tr}\,}

\begin{document}

\title{A noise-limiting quantum algorithm using mid-circuit measurements\texorpdfstring{\\}{Lg} for dynamical correlations at infinite temperature}

\author{Etienne Granet and Henrik Dreyer}
\affiliation{Quantinuum, Leopoldstrasse 180, 80804 Munich, Germany}
\date{\today}

\begin{abstract}

It is generally considered that the signal output by a quantum circuit is attenuated exponentially fast in the number of gates. This letter explores how algorithms using mid-circuit measurements and classical conditioning as \emph{computational tools} (and not as error mitigation or correction subroutines) can be naturally resilient to complete decoherence, and maintain quantum states with useful properties even for infinitely deep noisy circuits. Specifically, we introduce a quantum channel built out of mid-circuit measurements and feed-forward, that can be used to compute dynamical correlations at infinite temperature and canonical ensemble expectation values for any Hamiltonian. The unusual property of this algorithm is that in the presence of a depolarizing channel it still displays a meaningful, non-zero signal in the large depth limit. We showcase the noise resilience of this quantum channel on  Quantinuum's H1-1 ion-trap quantum computer.

\end{abstract}

\maketitle

\textbf{\emph{Introduction.}}--- The main obstacle to large-scale quantum computing is the imperfect realization of quantum operations \cite{preskill2018quantum,knill2005quantum}. In the presence of noise, the signal output by a quantum circuit is generically exponentially attenuated with the number of gates \cite{emerson2005scalable,moses2023race,mayer2021theory,proctor2022measuring,baldwin2022re}, causing exponential overheads which threaten the prospects of the utility of quantum computers.
Quantum error correcting codes are particular circuits fundamentally relying on mid-circuit measurements and classical conditioning on the measurement outcomes (also called \emph{feed-forward}) that delay this complete depolarizing for very deep circuits \cite{shor1995scheme,knill1998resilient,calderbank1996good,steane1996error,kitaev2003fault,nielsen2010quantum}. However, these circuits only serve as subroutines within unitary circuits and do not perform a computational task themselves. The purpose of this work is to investigate whether algorithms that use mid-circuit measurements and feed-forward directly as computational tools (rather than error correcting subroutines) are naturally \emph{noise-resilient} in the sense that in the limit of an infinitely deep circuit there remains a non-zero signal that contains information on the task fulfilled by the algorithm.

In this Letter, we answer this question in the affirmative. Specifically, we introduce a simple ``shuffling" quantum channel based on mid-circuit measurements and feed-forward that outputs a non-zero signal when repeated an arbitrary number of times, even in the presence of noise. Crucially, this non-zero signal also carries non-trivial, physically interesting information: We show that this quantum channel can be used to compute finite temperature expectation values for any Hamiltonian $H$, as well as dynamical correlations at infinite temperature $\tr[S e^{iHt}S e^{-iHt}]$ where $S$ is any Pauli string. We also show that it can prepare certain density matrices in a noise-resilient way as fixed points of a quantum channel.


The use of mid-circuit measurements to perform non-unitary operations is well known, for example as a means of post-selecting particular measurement outcomes \cite{kitaev1995quantum,harrow2009quantum,childs2012hamiltonian,berry2015simulating,berry2015hamiltonian,ge2019faster,schlimgen2022quantum,lee2020projected,choi2021rodeo,aaronson2005quantum}, to prepare specific, analytically tractable quantum states more efficiently \cite{iqbal2023creation,tantivasadakarn2022shortest,tantivasadakarn2021long,tantivasadakarn2023hierarchy, foss-feig2023demonstration, iqbal2023order}, or on top of another algorithm to mitigate the effect of noise \cite{mcardle2019error,botelho2022error,bonet2018low}. Some algorithms use a weak form of feed-forward where measurement outcomes are used as a stopping criterion or to partially reset the state \cite{cubitt2023dissipative,verstraete2009quantum,kraus2008preparation,xu2014demon,decross2022qubit,haghshenas2023probing}. The novelty in the present approach is the use of mid-circuit measurement and feed-forward as \emph{computational} primitives to obtain non-trivial physical properties which, as we will show, endows the computation with a natural noise-limiting property. This noise-limiting behaviour is demonstrated analytically with specific noise models, numerically with noisy simulations, and experimentally on Quantinuum ion-trap quantum computers.

\textbf{\emph{The shuffling quantum channel.}}--- Let us start with introducing the main quantum channel that we will study in this Letter. The effect of the channel on expectation values is given by Eq \eqref{result} below, and we will show that it can be used to compute dynamical correlations at infinite temperature. As we will explain, the surprising property of this channel is that when repeated several times in the presence of noise, it still outputs meaningful non-zero signal.

For a generic Hamiltonian $H$ on $L$ qubits, we define the quantum channel \texttt{Shuffle}$(H)$ by the three following steps.

\begin{enumerate}
    \item Pick a Pauli matrix at each site, i.e. $M_1,...,M_L\in \{I,X,Y,Z\}$ uniformly at random.
    \item Prepare an ancilla in state $|0\rangle_a$. On the $L$ qubits and the ancilla, apply $e^{iH_{M_1...M_L}X_a}$ where we defined
    \begin{equation}
        H_{M_1...M_L}=M_1...M_L H M_L...M_1\,.
    \end{equation}
    \item Measure the ancilla. Only if the measurement outcome is $0$, apply $M_1...M_L$ on the $L$ qubits.
\end{enumerate}
Let us denote $\rho$ the initial density matrix of the $L$ qubits, and $\rho'$ the new density matrix after running \texttt{Shuffle}$(H)$. We are going to show that for any Pauli string operator $S=M'_1...M'_L$ with $M'_i\in\{I,X,Y,Z \}$, we have
\begin{equation}\label{result}
    \tr[S \rho']=A \tr[S\rho]+B\,,
\end{equation}
with
\begin{equation}\label{AB}
    A=\frac{\tr[S\sin(H) S \sin(H)]}{2^L}\,,\qquad  B=\frac{\tr[S\cos^2(H)]}{2^L}\,.
\end{equation}
Moreover, the probability $p$ of measuring $0$ in the ancilla is
\begin{equation}\label{ancilla}
    p=\frac{\tr[\cos^2(H)]}{2^L}\,.
\end{equation}

\textbf{\emph{Proof.}}--- From step (1), we fix some Pauli matrices at each site $M_1,...,M_L\in\{I,X,Y,Z\}$. We will denote $\pmb{M}=M_1...M_L$ and $\mathcal{M}$ the set of all the $4^L$ Pauli strings. The operator $U=e^{i H_{\pmb{M}} X_a}$ is diagonal in the eigenbasis of $H_{\pmb{M}}$, with matrix elements $U_n$ that are operators on the ancilla and that read
\begin{equation}
   U_n= \left(\begin{matrix}
       \cos E_n & i\sin E_n\\
       i\sin E_n & \cos E_n
   \end{matrix} \right)\,,
\end{equation}
with $E_n$ the corresponding eigenvalue of $H_{\pmb{M}}$. By preparing the ancilla in the state $|0\rangle$, applying $U$ and measuring the ancilla, we act on the $L$ qubits with the operator $\langle 0| U |0\rangle=\cos(H_{\pmb{M}})$ if we measure $0$, and with $\langle 1| U |0\rangle=i\sin(H_{\pmb{M}})$ if we measure $1$. It follows that the density matrix $\rho'_{\pmb{M}}$ that we obtain after the steps (2) and (3), assuming $M_1,...,M_L$ have been picked at step (1), is
\begin{equation}\label{rhoprime}
\begin{aligned}
    \rho'_{\pmb{M}}&=\sin(H_{\pmb{M}})\rho \sin(H_{\pmb{M}})+\pmb{M}\cos(H_{\pmb{M}})\rho \cos(H_{\pmb{M}})\pmb{M}\\
    &=\sin(H_{\pmb{M}})\rho \sin(H_{\pmb{M}})+\cos(H)\pmb{M}\rho \pmb{M}\cos(H)\,.
\end{aligned}
\end{equation}

Taking into account all the $4^L$ possibilities of $M_1,...,M_L$ at step (1), the density matrix $\rho'$ after one run of \texttt{Shuffle}$(H)$ is the sum of all the $\rho'_{\pmb{M}}$'s divided by $4^L$. We now note that on the second term of $\rho'_{\pmb{M}}$, this sum precisely implements a completely depolarizing channel \cite{nielsen2010quantum}. Namely, for any density matrix $\rho$ we have
\begin{equation}\label{depolarizing}
   \frac{1}{4^L} \sum_{\pmb{M}\in\mathcal{M}} \pmb{M}\rho \pmb{M}=\frac{1}{2^L}\,.
\end{equation}
It follows
\begin{equation}\begin{aligned}
    \label{recurrence}
    \rho'=&\frac{1}{4^L}\sum_{\substack{\pmb{M}\in\mathcal{M}}} \sin(H_{\pmb{M}})\rho \sin(H_{\pmb{M}})+\frac{\cos^2(H)}{2^L}\,.
\end{aligned}
\end{equation}
We note that the feed-forward in step (3) is crucial to obtain a non-trivial density matrix. Had we always applied $\pmb{M}$ on the qubits independently of the measurement outcome of step (3), the density matrix $\rho'$ would have been the completely depolarized state $1/2^L$.

Let us now consider a given Pauli string operator $\pmb{M'}\in\mathcal{M}$, and denote the expectation values
\begin{equation}
    m=\tr[\pmb{M'}\rho]\,,\qquad m'=\tr[\pmb{M'}\rho']\,.
\end{equation}
Multiplying \eqref{recurrence} by $\pmb{M'}$ and taking the trace, we make appear in the sum the terms $\tr[\pmb{M'}\sin(H_{\pmb{M}})\rho \sin(H_{\pmb{M}})]$. Decomposing $\rho$ in the basis of Pauli strings
\begin{equation}
    \rho=\sum_{\pmb{M''}\in\mathcal{M}} c_{\pmb{M''}}\pmb{M''}\,,
\end{equation}
we write
\begin{equation}
\begin{aligned}
    &\tr[\pmb{M'}\sin(H_{\pmb{M}})\rho \sin(H_{\pmb{M}})]\\
    &=\sum_{\pmb{M''}\in\mathcal{M}} c_{\pmb{M''}}\tr[\pmb{M}\pmb{M'}\pmb{M}\sin(H)\pmb{M}\pmb{M''}\pmb{M} \sin(H)]\,.
\end{aligned}
\end{equation}
Let us perform first the sum over $M_i\in\{I,X,Y,Z \}$ appearing in \eqref{recurrence}. We note that $M_i M_i' M_i$ is either $M_i'$ (if $M_i=I$, or $M_i'=I$, or $M_i=M_i'$), or $-M_i'$ (in the remaining cases). If $M_i'\neq M_i''$, then at least one of the two is not $I$, say $M_i'\neq I$. If $M_i''= I$, then one sees that one gets an amplitude $\tr[\pmb{M'}\sin(H)\pmb{M''}\sin(H)]$ with a $+$ sign when $M_i=I$ and $M_i=M_i'$, and with a $-$ sign in the two remaining cases. Hence this term vanishes once summed over $M_i$. If $M_i'=I$ and $M''_i\neq I$, the same reasoning applies. If both $M'_i,M_i''\neq I$, one sees that one gets a $+$ sign when $M_i=I$ and $M_i\neq M_i',M_i''$, and a $-$ sign in the two remaining cases. Hence this term also vanishes once summed over $M_i$. It follows that once we sum over all the $M_i$, only the terms where $M_i'=M_i''$ give a non-zero contribution. This yields
\begin{equation}
\begin{aligned}
    &\sum_{\substack{\pmb{M}\in\mathcal{M}}} \tr[\pmb{M}\pmb{M'}\pmb{M}\sin(H)\pmb{M}\pmb{M''}\pmb{M}\sin(H)]\\
    &=4^L \tr[\pmb{M'}\sin(H)\pmb{M''}\sin(H)]\prod_{i=1}^L \delta_{M_i',M_i''}\,.
\end{aligned}
\end{equation}
Noting that we have $m=c_{\pmb{M'}}2^L$, we obtain thus
\begin{equation}\label{mprime}
\begin{aligned}
    m'=&m \frac{\tr[\pmb{M'}\sin(H)\pmb{M'}\sin(H)]}{2^L}+\frac{\tr[\pmb{M'}\cos^2(H)]}{2^L}\,,
\end{aligned}
\end{equation}
which is precisely \eqref{result}. To compute the probability $p$ of measuring $0$ in the ancilla, we note that if $M_1,...,M_L$ has been picked at step (1), then this probability $p_{\pmb{M}}$ is
\begin{equation}
    p_{\pmb{M}}=\tr[\rho \cos^2(H_{\pmb{M}})]\,.
\end{equation}
Summing over $M_1,...,M_L$, this yields
\begin{equation}
\begin{aligned}
    p=&\frac{\tr[\cos^2(H)]}{4^L}\times\sum_{\substack{\pmb{M}\in\mathcal{M}}}\tr[\rho \pmb{M} \frac{\cos^2(H)}{\tr[\cos^2(H)]}\pmb{M}]\,.
\end{aligned}
\end{equation}
Using in this equation the relation \eqref{depolarizing} for the density matrix $\frac{\cos^2(H)}{\tr[\cos^2(H)]}$\, we obtain \eqref{ancilla} indeed.

\textbf{\emph{Dynamical correlations at infinite temperature.}}--- Let us show how the shuffling quantum channel can be used to compute dynamical correlations of Pauli strings $S=M_1...M_L$ at infinite temperature, defined as
\begin{equation}
    C_S(t)=\frac{\tr[e^{-iHt}S e^{iHt}S ]}{2^L}\,.
\end{equation}
These quantities are relevant for transport properties and have attracted particular interest recently \cite{rosenberg2023dynamics,gopalakrishnan2019anomalous,de2020superdiffusion,kanasz2017quantum,scheie2021detection,granet2020systematic,granet2022out,rakovszky2022dissipation,vznidarivc2011spin,dupont2020universal,parker2019universal}. 
To compute this quantity with the shuffling quantum channel, we write
\begin{equation}\label{cst}
    C_S(t)=\frac{\tr[S \cos(Ht)S\cos(Ht)]}{2^L}+\frac{\tr[S \sin(Ht)S\sin(Ht)]}{2^L}\,.
\end{equation}
Let us consider $\rho$ an arbitrary initial density matrix for which we know $\tr[\rho S]$. We then apply on $\rho$ either \texttt{Shuffle}$(Ht)$ with probability $1/2$, or \texttt{Shuffle}$(Ht+\tfrac{\pi}{2})$ with probability $1/2$. We measure the Pauli string $S$ and denote $\tr[\rho' S]$ the average outcome. Using formula \eqref{result}, we have thus
\begin{equation}
    \tr[\rho' S]=\frac{A_0+A_{\pi/2}}{2}\tr[\rho S]+\frac{B_0+B_{\pi/2}}{2}\,,
\end{equation}
with $A_x,B_x$ the coefficients \eqref{AB} corresponding to the Hamiltonian $Ht+x$. From \eqref{AB} we find thus for $S\neq I$
\begin{equation}\label{rhoprime2}
    \tr[\rho' S]=\frac{C_S(t)\tr[\rho S]}{2}\,.
\end{equation}
Hence, one can directly compute the value of $C_S(t)$ provided we chose $\rho$ such that $\tr[\rho S]\neq 0$. We note there are other similar ways of extracting $C_S(t)$ from this quantum channel \cite{supp}.

This algorithm for dynamical correlations at infinite temperature presents advantages compared to previously existing techniques. Computing the trace over the Hilbert space through a purification requires to double the number of qubits, whereas here there is only one ancilla qubit overhead. Implementing the trace through a Haar random state preparation \cite{summer2023calculating} has a gate overhead to prepare the Haar random state, and also a shot overhead because it computes the square of the dynamical correlations and can only treat one site at a time, which increases considerably the variance. Finally, for $S$ single Pauli matrices, evaluating the trace as a sum over product states in the basis of $S$ can have a large statistical sampling overhead if dynamical correlations are small in this basis, whereas our algorithm is basis-independent.

\textbf{\emph{Mixed state preparation.}}--- The density matrix constructed previously in \eqref{rhoprime2} contains information about the physics of the system through $C_S(t)$, but also on the initial density matrix through $\tr[\rho S]$. This can be problematic for strings $S$ involving different Pauli matrices at different sites, since the preparation of $\rho$ such that $\tr[\rho S]\neq 0$ for multiple such $S$ would become non-trivial.

The shuffling quantum channel through the relation \eqref{result} actually enables one to prepare non-trivial density matrices  independently of the initial density matrix. Let us repeat $n$ times the channel \texttt{Shuffle}$(H)$ on an initial density matrix $\rho_0$, denoting $\rho_n$ the resulting density matrix. Eq \eqref{result} yields a geometric series for any string $S$
\begin{equation}
    \tr[S\rho_n]=A^n \tr[S\rho_0]+\frac{1-A^n}{1-A}B\,,
\end{equation}
with $A,B$ given in \eqref{AB}. In particular, since $|A|<1$ (except for very specific choices of $H,S$ where we would have $|A|=1$) when $n\to\infty$ we obtain
\begin{equation}
    \underset{n\to\infty}{\lim}\, \tr[S\rho_n]=\frac{B}{1-A}\,,
\end{equation}
which does not depend on the initial density matrix. Moreover the convergence is exponentially fast in $n$ with rate $\log 1/|A|$. We thus prepare a non-trivial density matrix whose expectation values contain physical information about the system studied.

\textbf{\emph{Noise-limiting behaviour: analytics.}}--- We now argue that the algorithms based on the repetition of the shuffling quantum channel are to some extent resilient to noise. Let us consider a noise model where a global depolarizing channel with amplitude $\lambda$ is applied on the $L$ qubits every time we apply the \texttt{Shuffle}$(H)$ channel. This depolarizing channel transforms a density matrix $\rho$ into
\begin{equation}
    \rho\mapsto (1-\lambda)\rho+\frac{\lambda}{2^L}\,.
\end{equation}
Hence, using \eqref{result}, the (noisy) density matrix $\rho'_{\rm noisy}$ that describes the $L$ qubits after one round of \texttt{Shuffle}$(H)$ and one global depolarizing channel satisfies for $S\neq I$
\begin{equation}
    \tr[S \rho'_{\rm noisy}]=(1-\lambda)(A \tr[S\rho]+B)\,.
\end{equation}
Denoting $\rho_{{\rm noisy},n}$ the density matrix obtained after $n$ noisy applications of \texttt{Shuffle}$(H)$, the limiting value reached in the limit of an infinite number of applications is
\begin{equation}\label{B1-Anoisy}
    \underset{n\to\infty}{\lim}  \tr[S\rho_{{\rm noisy},n}]=\frac{(1-\lambda)B}{1-(1-\lambda)A}\,,
\end{equation}
which is \emph{finite}, even if an infinite number of gates and noise channels have been applied. One notes that none of the $L$ qubits are ever measured or re-initialized. This shows the noise resilience of the preparation of density matrices through the shuffling quantum channel.

Let us explain why this noise-limiting behaviour is directly related to the use of mid-circuit measurements and feed-forward. A circuit involving only unitary operations and measurements without feed-forward can always be written as the composition of density matrix maps $T(\rho)=U\rho U^\dagger$ with $U$ unitary or $T(\rho)=\sum_m P_m \rho P_m^\dagger$ with $\sum_m P_m  P_m^\dagger={\rm Id}$. These maps both satisfy $T({\rm Id})={\rm Id}$ (also called ``unital" maps), so the completely mixed state $\frac{{\rm Id}}{2^L}$ is always a fixed point of such circuits. However, with feed-forward these density matrix maps are in general $T(\rho)=\sum_m U_m P_m \rho P_m^\dagger U_m^\dagger$ with $U_m$ unitary, for which the identity is (generically) not a fixed point. This prevents thus the completely depolarized state to be reached in the limit of large number of gates. 

This explanation yields the following nice interpretation of the noise resilience of our algorithm. The resulting density matrix after a large number of rounds converges to the fixed point of the quantum channel. Because of the presence of mid-circuit measurements and feed-forward, this fixed point cannot be (in general) the completely depolarized channel. Now, incorporating noise in the process (under any form) will only smoothly change the quantum channel, and so only smoothly change the fixed point, instead of sending it to the completely depolarized state. The noise-resilience of the prepared density matrix is thus built into the algorithm itself.

\begin{figure}[tb]
    \centering
    \includegraphics[scale=0.4]{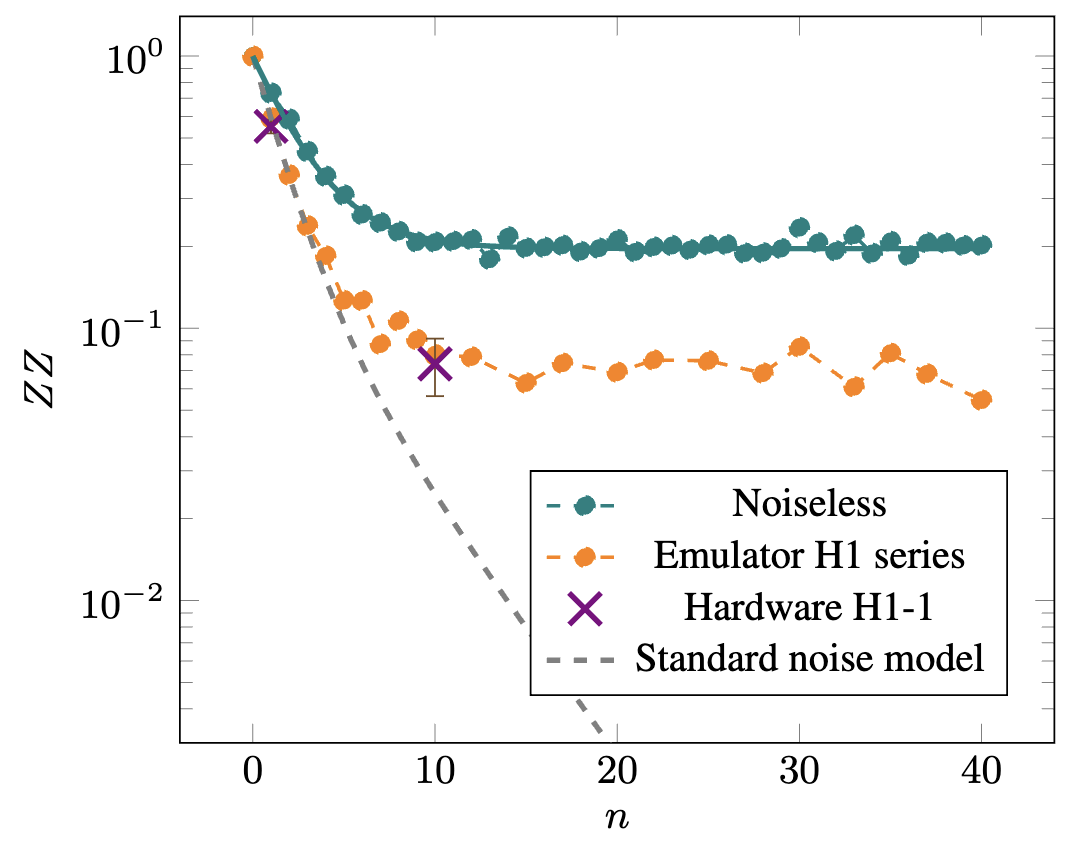}
    \caption{\emph{Noise-limiting property of the shuffling quantum channel}. Expectation value $ZZ\equiv\frac{1}{L}\sum_{i=1}^L\langle Z_iZ_{i+1}\rangle$ as a function of the number $n$ of calls to \texttt{Shuffle}$(H)$, with $H$ given by \eqref{H}, and using $5$ Trotter steps to implement $e^{iH}$. There are $1000$ shots per circuit for simulations, and $500$ shots per circuit for hardware. The errors bars indicate $\pm$ one standard deviation. After compilation, there are $185n$ two-qubit gates and $782n$ one-qubit gates as a function of $n$.}
    \label{noise}
\end{figure}

\begin{figure}[tb]
    \centering
    \includegraphics[scale=0.4]{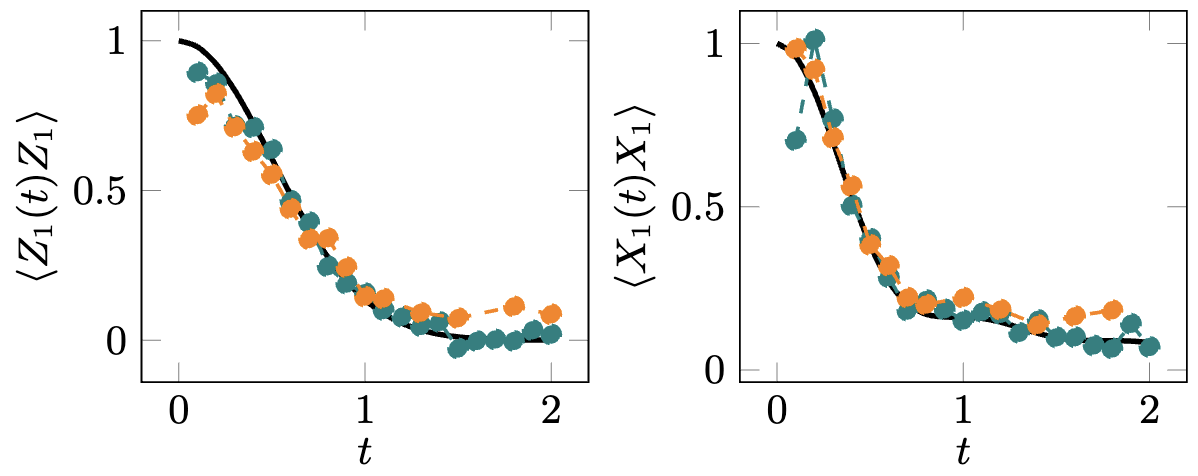}
    \caption{\emph{Dynamical correlations at infinite temperature}. Expectation values $\tr[Z_1(0)Z_1(t)]/2^L$ (left panel) and $\tr[X_1(0)X_1(t)]/2^L$ (right panel) computed with the shuffling quantum channel with the technique described in the Supplemental Material (SM) \cite{supp}, as a function of $t$, with $G(t)=e^{iHt}G e^{-itH}$ where $H=H_{\rm Ising}$ on $L=8$ sites: noiseless simulations (teal), noisy emulator H1 series (orange) and exact value (black, continuous line). We use Trotter steps $dt=0.1$ to implement the time evolution. There are $1000$ shots per circuit. No noise mitigation technique is used.}
    \label{dyna}
\end{figure}

\textbf{\emph{Noise-limiting behaviour: experiments and simulations.}}--- We now present numerical and experimental evidence of the noise resilience of our algorithm. We consider the 1D Ising model in a transverse field $h=1$ and $t=0.1$ on $L=8$ sites with periodic boundary conditions
\begin{equation}\label{H}
  H= \frac{\pi}{3}+t H_{\rm Ising}\,,\quad H_{\rm Ising}=-\sum_{j=1}^L Z_jZ_{j+1}-h\sum_{j=1}^L X_j\,.
\end{equation}
We first focus on the shuffling quantum channel \texttt{Shuffle}$(H)$. We initialize the $L$ qubits in $|0\rangle$ and measure $\frac{1}{L}\sum_{i=1}^L Z_iZ_{i+1}$ after $n$ rounds of \texttt{Shuffle}$(H)$. We chose this observable as $\frac{1}{L}\sum_{i=1}^L X_i$ or $\frac{1}{L}\sum_{i=1}^L Z_i$ would have a  zero expectation value at either small or large number of rounds. We report in Fig \ref{noise} the results of noiseless simulations, noisy simulations with the H1 emulator, and actual hardware implementation with the H1-1 machine of Quantinuum for number of rounds $n=1$ ($185$ two-qubit gates and $782$ one-qubit gates) and $n=10$ ($1850$ two-qubit gates and $7820$ one-qubit gates). It is known that noise on quantum computers can be well accounted for, in a first approximation, by attenuating the signal with a factor equal to the product of the fidelities of the gates entering the circuit. We  evaluate the effect of the noise by measuring the signal after one round $n=1$ on the emulator with a large number of shots, and comparing it to the exact value. From this, we obtain an attenuation per round $1-\lambda$. We plot then in Fig \ref{noise} the effect of the noise assuming it can be modelled by an attenuation factor $(1-\lambda)^n$ at round $n$ times the exact result, which is labelled by ``standard noise model". 

We observe that the results from the hardware agree very well with the emulator data. In particular the measured value at $n=10$ is incompatible with the standard noise model by around $3$ standard deviations. If we used the hardware data at $n=1$ to calibrate the standard noise model instead of the emulator data, the incompatibility would be even larger. 

Secondly, in Fig \ref{dyna} we implement our algorithm for dynamical correlations at infinite temperature, comparing the exact values, the noiseless simulations and the results of the Quantinuum hardware emulator, without doing any noise mitigation. This emulator implements in particular a two-qubit gate fidelity of around $0.998$. We observe that the noisy values agree very well with the noiseless, despite the circuits containing several hundreds of two-qubit gates, showing a noise-resilience. 

\textbf{\emph{Canonical ensemble expectation values.}}--- Finally, we show that using a shuffling quantum channel with multiple ancillas, one can compute expectation values at finite temperature. Although this application is not of practical interest (as we will show, it is exponential in system size runtime), it shows the versatility of our channel in extracting physical information about a system. Let us consider a Hamiltonian $H$ on $L$ qubits, an observable $O$ and an inverse temperature $\beta$. We define
\begin{equation}\label{Hprime}
    H'=\arctan \theta+\frac{\beta H-\epsilon Z_v O}{2\theta N}\,,
\end{equation}
on the system of $L$ qubits plus one ``virtual" qubit $v$, where $Z_v$ denotes the Pauli matrix $Z$ applied on the virtual qubit $v$, and where we introduced parameters $\epsilon>0$ and $\theta>0$, as well as an integer $N$. As shown in the SM \cite{supp}, the shuffling quantum channel can be generalized into a channel \texttt{Shuffle}$(H,N)$ that uses $N$ ancillas as follows: one applies step (2) to each of the $N$ ancillas, and then in step (3) one applies $M_1...M_L$ on the $L$ qubits only if \emph{all} the $N$ ancillas have been measured to be in state $0$. An identical relation to \eqref{result} follows then with precise coefficients $A$ and $B$ \cite{supp}. In the particular case of $H'$ in \eqref{Hprime}, because $Z_v$ commutes with $H'$, we have
\begin{equation}
    \tr[\rho' Z_v]=A\tr[\rho Z_v]+B\,,
\end{equation}
with
\begin{equation}\label{a7}
   A=1-\frac{\Tr[\cos^{2N}(H')]}{2^{L+1}}\,,\quad B=\frac{\Tr[Z_v\cos^{2N}(H')]}{2^{L+1}}\,,
\end{equation}
where $\Tr$ denotes the trace on the $L$ qubits and the virtual ancilla $v$. Hence, by repeating this generalized channel we obtain
\begin{equation}\label{ratio}
    \underset{n\to\infty}{\lim}  \Tr[Z_v \rho_n]=\frac{\Tr[Z_v\cos^{2N}(H')]}{\Tr[\cos^{2N}(H')]}\,.
\end{equation}
This prepares the virtual qubit in an ensemble that is close to a canonical ensemble. Indeed, for $N$ larger than the norm of $\beta H/\theta$, we have \cite{lu2021algorithms}
\begin{equation}\label{canoni}
    \cos^{2N}(H')\sim \frac{e^{-\beta H+\epsilon Z_v O}}{(1+\theta^2)^N}\,.
\end{equation}
Hence, performing the trace over the virtual qubit $v$ we obtain in this regime
\begin{equation}\label{tracezv}
    \underset{n\to\infty}{\lim}  \Tr[Z_v \rho_n]=\epsilon\frac{\tr[O e^{-\beta H}]}{\tr[e^{-\beta H}]}+\mathcal{O}(\epsilon^2, \tfrac{\beta||H||}{N\theta})\,,
\end{equation}
where we now use $\tr$ to denote the trace over the $L$ qubits, without the virtual qubit $v$. Interestingly, the algorithm automatically computes the \emph{ratio} of traces $\tr[O e^{-\beta H}]/2^L$ and $\tr[e^{-\beta H}]/2^L$ that can each of them be exponentially small. Using \eqref{a7} and \eqref{canoni}, the number of rounds $n$ required to reach this limit is $1/\log (1-\frac{\tr[e^{-\beta H}]}{2^{L}(1+\theta^2)^N})$. The time complexity is thus exponential in the system size $L$, which is the expected scaling for a finite-temperature algorithm that applies to generic Hamiltonians.

\textbf{\emph{Discussion.}}--- 
In this Letter we have provided a first example of how non-error-corrected algorithms can avoid total depolarisation in the infinite-depth limit. This proof of principle opens up a number of possible future directions. For example, the algorithms presented here for finite temperature expectation values have an exponential time complexity, which is unavoidable without further assumption on $H$. It would be very valuable to investigate whether these features can be improved for local Hamiltonians $H$.

We acknowledge helpful discussions with Eli Chertkov, Matthew DeCross, Kevin Hemery, Michael Foss-Feig, Ramil Nigmatullin, Reza Hagshenas and Charlie Baldwin. E.G. acknowledges support by the Bavarian Ministry of Economic Affairs, Regional Development and Energy (StMWi) under project Bench-QC (DIK0425/01). The emulator data in this work was produced in June-July 2023.
The experimental data in this work was
produced by the Quantinuum H1-1 trapped ion quantum
computer, Powered by Honeywell, in July 2023.

\bibliography{bibliography}

\newpage

\onecolumngrid



\section{-SUPPLEMENTAL MATERIAL-}

\section{Generalization of the channel to multiple ancillas.}
In this Section we generalize the shuffling quantum channel to multiple ancillas. Namely, for an integer $N$ we define \texttt{Shuffle}$(H,N)$ through the following steps.

\begin{enumerate}
    \item Pick a Pauli matrix at each site, i.e. $M_1,...,M_L\in \{I,X,Y,Z\}$ uniformly at random.
    \item Prepare $N$ ancillas in state $|0\rangle_{a_1}... |0\rangle_{a_N}$. For $\ell=1,...,N$, apply $e^{iH_{M_1...M_L}X_{a_\ell}}$.
    \item Measure the $N$ ancillas. Only if the measurement outcome is $0$ for all the $N$ ancillas, apply $M_1...M_L$ on the $L$ qubits.
\end{enumerate}
The analysis of \texttt{Shuffle}$(H)$ is straightforwardly generalized to the case $N>1$. For any Pauli string operator $S=M'_1...M'_L$ with $M'_i\in\{I,X,Y,Z \}$, we have
\begin{equation}\label{result2}
    \tr[S \rho']=A \tr[S\rho]+B\,,
\end{equation}
with
\begin{equation}\label{A2}
\begin{aligned}
    A=\frac{1}{2^L}\sum_{p=0}^{N-1}\binom{N}{p} \tr[&S \cos^p(H) \sin^{N-p}(H)\\
    &\times S \cos^p(H) \sin^{N-p}(H)]\,,
\end{aligned}
\end{equation}
and
\begin{equation}\label{B2}
    B=\frac{\tr[S\cos^{2N}(H)]}{2^L}\,.
\end{equation}
Moreover, the probability $p$ of measuring $0$ in all the $N$ ancillas is
\begin{equation}\label{ancilla2}
    p=\frac{\tr[\cos^{2N}(H)]}{2^L}\,.
\end{equation}
When $S$ commutes with $H$, we obtain
\begin{equation}
    A=1-\frac{\tr[\cos^{2N}(H)]}{2^L}\,.
\end{equation}
In the main text, the system on which we apply this quantum channel contains $L$ qubits and one extra ancilla $v$. It is thus a system with $L+1$ qubits. Moreover the Hamiltonian $H'$ commutes with the particular choice of Pauli string $S=Z_v$ that we consider. This yields thus formula (28) in the main text.

\section{Alternative way of extracting the coefficients $A$ and $B$.}

In this Section we present an alternative way of computing coefficients $A$ and $B$ appearing in the shuffling quantum channel output in Eq (2) of the main text. By applying twice \texttt{Shuffle}$(H)$ on a density matrix $\rho$, and denoting $\rho''$ the output density matrix, we have for any Pauli string $S$
\begin{equation}
    \tr[\rho'' S]=A^2\tr[\rho S]+AB+B\,.
\end{equation}
It follows thus that we can extract $A$ from the ratio
\begin{equation}\label{deduceA2}
    A=\frac{\tr[\rho'' S]-\tr[\rho' S]}{\tr[\rho' S]-\tr[\rho S]}\,,
\end{equation}
where we recall that $\rho'$ denotes the density matrix after one single run of \texttt{Shuffle}$(H)$. Then, $B$ is given by
\begin{equation}
    B=\tr[\rho' S]-A\tr[\rho S]\,.
\end{equation}
 Applying this process to the Hamiltonians $Ht$ and $Ht+\tfrac{\pi}{2}$, one can then sum the resulting $A$'s and obtain the dynamical correlation $C_S(t)$. This is the technique we used in Fig 2 of the main text to compute dynamical correlations at infinite temperature.

This way of computing $C_S(t)$ presents sometimes a lower variance compared to the one quoted in the main text, although more involved. Let us indeed compute the variance $\Delta$ in the estimation of $A$. The variance in estimating $\tr[\rho' S]$ using $K$ shots is $\frac{1-\tr[\rho' S]^2}{K}$. Hence, using Gaussian error propagation, at large $K$ the variance $\Delta$ in estimating $A$ through formula \eqref{deduceA2} is
\begin{equation}\begin{aligned}
    \Delta=&\frac{1-\tr[\rho' S]^2}{K}\frac{1}{(\tr[\rho S]-\tr[\rho' S])^2}+\frac{1-\tr[\rho'' S]^2}{K}\frac{1}{(\tr[\rho S]-\tr[\rho' S])^2}\\
    &+\frac{1-\tr[\rho S]^2}{K}\left( \frac{\tr[\rho' S]-\tr[\rho'' S]}{(\tr[\rho S]-\tr[\rho' S])^2}\right)^2+\frac{1-\tr[\rho' S]^2}{K}\left( \frac{\tr[\rho' S]-\tr[\rho'' S]}{(\tr[\rho S]-\tr[\rho' S])^2}\right)^2\,.
\end{aligned}
\end{equation}
Expressed in terms of $m\equiv \tr[\rho S]$ and $A,B$, this is
\begin{equation}
    \begin{aligned}
        \Delta=\frac{1}{K(m(A-1)+B)^2}\left[2+2A^2-(mA+B)^2-(mA^2+AB+B)^2-m^2A^2-(mA+B)^2A^2 \right]\,.
    \end{aligned}
\end{equation}
Dynamical correlations typically become small as $t$ increases, meaning that $A$ is typically small, whereas $B$ remains of order $1$ since the two $B$'s obtained from $Ht$ and $Ht+\tfrac{\pi}{2}$ have to add up to $1$. When $A\to 0$ and $B\to 1$, we see that we have $\Delta\to 0$ at fixed $K$. When $A\to 0$ for $B=1/2$ we have $\Delta=\frac{2}{3K}$ if we choose $m=-1$. For the method stated in the text, the variance in estimating $C_S(t)$ is $\frac{4}{K}$ when $C_S(t)\to 0$. This approach yields thus typically a lower variance in these situations.

\end{document}